\documentclass[10pt, conference]{IEEEtran}
\IEEEoverridecommandlockouts
\usepackage{cite}
\usepackage{amsmath,amssymb,amsfonts}
\usepackage{algorithmic}
\usepackage{graphicx}
\usepackage{textcomp}
\usepackage{xcolor}
\def\BibTeX{{\rm B\kern-.05em{\sc i\kern-.025em b}\kern-.08em
    T\kern-.1667em\lower.7ex\hbox{E}\kern-.125emX}}

\usepackage{tabularx}
\usepackage{cite}
\usepackage{amsmath,amssymb,amsfonts}
\usepackage{algorithm}
\usepackage{algorithmic}
\usepackage{graphicx}
\usepackage{textcomp}
\usepackage{xcolor}
\usepackage{comment}
\usepackage{subcaption}
\usepackage{tikz}
\usepackage{pgfplots}
\pgfplotsset{compat=1.17}
\usepackage{amsmath}
\usepackage{multicol}
\usepackage{booktabs} % \toprule, \midrule, \cmidrule, \bottomrule
\usepackage{multirow} % if you later want multi-row cells
\usepackage{adjustbox}
\usepackage{tcolorbox}
\usepackage{float}

\title{LLM Agents for Automated Dependency Upgrades}

\author
{\IEEEauthorblockN{Vali Tawosi}
\IEEEauthorblockA{
\textit{J.P. Morgan AI Research}\\
London, UK \\
vali.tawosi@jpmorgan.com}
\and
\IEEEauthorblockN{Salwa Alamir}
\IEEEauthorblockA{
\textit{J.P. Morgan AI Research}\\
London, UK \\
salwa.alamir@jpmchase.com}
\and
\IEEEauthorblockN{Xiaomo Liu}
\IEEEauthorblockA{
\textit{J.P. Morgan AI Research}\\
New York, USA \\
xiaomo.liu@jpmchase.com}
\and
\IEEEauthorblockN{Manuela Veloso}
\IEEEauthorblockA{
\textit{J.P. Morgan AI Research}\\
New York, USA \\
manuela.veloso@jpmchase.com}
}

\begin{document}

\maketitle

\begin{abstract}
    As a codebase expands over time, its library dependencies can become outdated and require updates to maintain innovation and security. However, updating a library can introduce breaking changes in the code, necessitating significant developer time for maintenance. To address this, we introduce a framework of LLM agents to be used in combination with migration documentation to automatically recommend and apply code updates and ensure compatibility with new versions. Our solution can automatically localize updated library usages in live Java codebases and implement recommended fixes in a user-friendly manner. The system architecture consists of multiple key components: a Summary Agent, Control Agent, and Code Agent. To validate our approach, we apply the framework on an industrial use case by which we create three synthetic code repositories with major Upgrade changes and benchmark our approach against state-of-the-art methods. Results show that our approach not only performs upgrades using fewer tokens across all cases but also achieves a precision of 71.4\%, highlighting its efficiency and effectiveness compared to state-of-the-art methods.
\end{abstract}


\section{Introduction}

Modern software projects strive to avoid reinventing the wheel by leveraging reusable functionality provided by open-source libraries. However, updating these libraries can introduce breaking changes in the code, posing challenges for developers \cite{KESHANI2023111738}. Many software projects heavily depend on open-source libraries that offer ready-to-use functionality, including those available for Java. For example, Maven Central is a repository for Java libraries, providing developers with access to a vast array of libraries for their projects \cite{Ede_2025}. While these libraries offer convenience and accessibility, they also introduce dependencies, and frequent library updates can result in breaks in developers' code. 

In the Java ecosystem, contributors often use warnings to alert library users of upcoming changes to the API and functionality. These warnings are typically issued several releases before the actual changes take effect, allowing developers time to update their code for compatibility \cite{zhong2024empiricalstudypackageleveldeprecation}. However, there is currently no clearly established method for automatically updating library usages in code to account for changes in Java libraries, leading to costly and time-consuming manual maintenance without immediate incentives. This issue is particularly pronounced in larger codebases, where developers might ignore warnings or lock into a specific library version \cite{sawant_2019, Kula_2017}. Such strategies can hinder a team's ability to innovate in the long term, preventing them from benefiting from new features, performance improvements, and bug fixes offered by updated libraries.

Some automated solutions have been proposed in the past, but have not been widely adopted, giving way to further research opportunities in this domain. Furthermore, large language models (LLMs) have been at the forefront of AI, and the software engineering domain is no exception. LLM systems have been used to tackle a number of coding challenges, with a focus on code generation and code completion \cite{chen_2021evaluating}, or automated program repair \cite{jimenez-2023-swebench}. Nevertheless, the maintenance and evolution of code remains a critical task in engineering, impacting code quality and reusability \cite{Raemaekers_2016}. As such, in this paper we propose a framework of LLM Agents for automated dependency upgrades. This system builds upon our previous work, the Autonomous LLM-based Multi-Agent Software Engineering Framework (ALMAS) \cite{tawosi2025almas}. The system is evaluated on an industrial use case with the aim to update code repositories after completing a major library upgrade.

\begin{figure*}[t!]
  \centering
  \includegraphics[width=\textwidth, trim={0 8.5cm 4cm 0},clip]{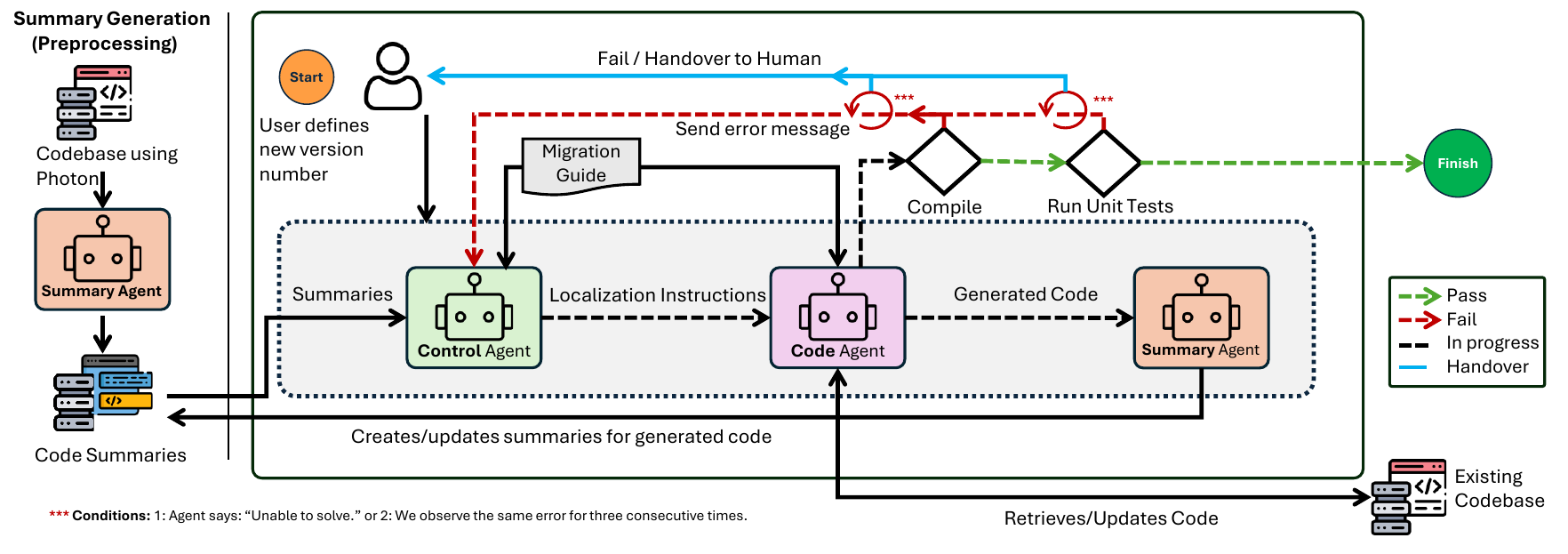}
  \caption{Diagram of the multi-agent LLM framework for automated Java dependency upgrades, showing the process of code summarization, change localization, and implementation, with iterative compilation and testing, and handover to a human.}
  \label{fig:architecture}
\end{figure*} 

\section{Related Work}

Automated codebase updates for dependency compatibility have been explored previously, often by analyzing differences between library versions. SemDiff \cite{Dagenais_2009} recommends client code adaptations by tracking framework changes, while LIBSYNC \cite{nguyen2010graph} learns API adaptation patterns from clients that have already migrated to newer versions.

Most prior work focuses on statically typed languages, though recent efforts address dynamic languages. For example, Nielsen et al. use semantic patches to locate breaking changes in JavaScript \cite{nielsen2021semantic}. Other approaches include manual changelog labeling, basic NLP for error-documentation matching \cite{Ansong2021soar}, and neural methods.

Neural machine translation (NMT) models, adapted from NLP, treat code tasks like bug fixing \cite{Jiang2021CURECN}, summarization \cite{Mastropaolo2021T5}, and review automation \cite{Hellendoorn2021codereview} as sequence-to-sequence problems using LSTM \cite{deepfix} \cite{Chen2021SequenceRSL} or Transformer architectures \cite{Chirkova2021empiricaltransformer}. However, source code generation faces challenges in syntax and semantics, prompting solutions like syntax-aware decoders \cite{zhu-etal-2013-fast} and AST-based methods \cite{Kim2021ast, alamir_2022, navarro_2023}.

Recent work leverages pre-trained code language models (CodeLMs) such as CodeBERT, CodeGPT, and CodeT5 for code updates. Liu et al. found that CodeLMs struggle with time-sensitive, complex updates and generalization \cite{liu2024automaticallyrecommendcodeupdates}. PCREQ, a six-component system, outperforms popular LLMs by 18–20\% \cite{lei2025pcreqautomatedinferencecompatible}. Multi-agent LLM systems have also shown promise in automated program repair on benchmarks like SWE-Bench \cite{yang2024sweagent}.

Our work frames code updates as sequential bug-fixing tasks, using a multi-agent LLM system \cite{tawosi2025almas} that utilises code summaries, iteratively updates code, and ensures successful builds and tests, with automatic handover to human intervention when needed.

\section{Methodology}

Our approach (depicted in Fig. \ref{fig:architecture}) leverages Meta-RAG, a change localisation mechanism designed for efficient code editing in large codebases by summarising code into structured natural language, reducing token count by nearly 80\% and enabling scalable information retrieval for localisation tasks \cite{tawosi2025meta}. This summarisation step provides a concise, coherent overview of the codebase, facilitating downstream editing and upgrades. The main workflow is built on a customised ALMAS multi-agent framework, which iteratively applies changes and resolves issues until the project builds and all tests pass, mirroring standard developer upgrade practices \cite{tawosi2025almas}.

\subsection{Preprocessing}
As a preprocessing step, the \textit{Summary Agent} generates and stores code summaries aligned with the code’s Abstract Syntax Tree (AST). Each code file and its units (functions/classes) receive one-line summaries describing their responsibilities, ensuring accurate mapping between code and summary structures \cite{tawosi2025meta}. These summaries are then used for retrieval and localisation, allowing RAG to operate on metadata rather than raw code. Summarisation is performed once for the entire codebase, with incremental updates as the code evolves.

\subsection{Main process}
The main process (depicted in Fig. \ref{fig:architecture} inside the gray box) makes use of a number of agents and knowledge sources to complete the upgrade. We describe the agents and their responsibilities before diving into the process workflow.

\subsubsection{Control Agent}
The Control Agent, central to Meta-RAG and ALMAS \cite{tawosi2025almas}, identifies code units to read (for context) and to write (for modification or creation) based on the current task. It first examines file-level summaries to select relevant files, then drills down to finer-grained summaries to pinpoint code units for context and editing. This enables efficient code retrieval and encourages code reuse and natural integration of changes.

\subsubsection{Code Agent}
The Code Agent (e.g., GPT-4o) receives the task and selected code units, then implements the required changes. By minimising context length, the agent avoids repeated code searches typical of other approaches. After editing, it triggers the Summary Agent to update summaries, maintaining alignment between code and metadata

\begin{table*}[h!]
\centering
\caption{Comparative results of the multi-agent LLM framework (LADU) against benchmark, showing the number of files, common files with Gold (G), lines added or removed, and common lines with (G) for each version upgrade scenario.}
\label{tab:results}

\begin{tabularx}{\textwidth}{l l l |l l l | l l l} % Specify column alignments here
\toprule
\textbf{Approach} & \textbf{Start} & \textbf{Target} & \textbf{\#Files} & \textbf{\#File (G)} & \textbf{\#Common Files} & \textbf{\#Lines A:R} & \textbf{\#Lines A:R (G)} &\textbf{\#Common Lines}  \\
\midrule
OpenHands + Claude 3.7 Sonnet & 3.1.12 & 3.2.11 & 8 & 7 & 7 & 33:78 &  38:62 & 14:45\\
OpenHands + Claude 3.7 Sonnet & 3.2.11 & 3.3.9 & 7 & 9 & 5 & 2:128 & 50:37 & 2:22\\ 
OpenHands + Claude 3.7 Sonnet & 3.3.9 & 3.4.2 & 7 & 7 & 6 & 6:132 & 23:31 & 4:9\\
\textbf{LADU + GPT-4o} & 3.1.12 & 3.2.11 & 8 & 7 & 7 & 37:67 & 38:62 & 13:48\\
\textbf{LADU + GPT-4o} & 3.2.11 & 3.3.9 & 3 & 9 & 3 & 4:7 & 50:37 & 2:5\\ 
\textbf{LADU + GPT-4o} & 3.3.9 & 3.4.2 & 3 & 7 & 3 & 10:9 & 23:31 & 4:5\\
\bottomrule
\end{tabularx}
\end{table*}

\begin{table}[h!]
\centering
\caption{Execution results of the multi-agent LLM framework (LADU) against benchmark.}
\label{tab:runtime}
\resizebox{\columnwidth}{!}{
\begin{tabular}{l | lll | lll} % Specify column alignments here
\toprule
\multicolumn{1}{l|}{\textbf{Metric}} & \multicolumn{3}{c|}{\textbf{OpenHands}} & \multicolumn{3}{c}{\textbf{LADU}}\\
%\cline{2-7}
 & 3.1 & 3.2 & 3.3 & 3.1 & 3.2 & 3.3 \\
\midrule
Number of steps & 106 & 86 & 72 & 18 & 16 & 4\\
Execution runtime [s] & 1020.0 & 720.0 & 450.0 & 473.6 & 424.0 & 277.4 \\
Total Tokens & 1,514,456 & 822,781 & 594,366 & 78,421 & 54,812 & 14,387 \\
Cost [\$] & 4.66 & 2.38 & 1.83 & 0.62 & 0.45 & 0.11\\
\bottomrule
\end{tabular}
}
\end{table}

\subsubsection{Process Workflow}

The upgrade process starts with user initiation, specifying a target version or retrieving it automatically from artifactory. The control agent accesses the codebase and generates a structure for every pom and yml file in the project, as these files are not included in the summary base.
The Control Agent also consults migration guides if available. It generates upgrade instructions, identifying files and changes needed. The Code Agent applies these changes, updating dependencies and configurations. The project is then compiled and tests ran; if errors occur, logs are sent back to the Control Agent for further resolution, following an Automated Program Repair (APR) loop. 
The cycle of applying changes, compiling the code, running tests, and reviewing the output log continues until the build and test run is successful. The code agent indicates its inability to resolve the issue or if the same error occurs more than $n$ times (with $n=3$), the run stops. These conditions are implemented to prevent an infinite loop that could exhaust resources.
The framework supports collaboration, allowing handover to human developers with a summary of actions taken (stored while the process was running), and resumption by AI agents if needed.

\section{Upgrades Use Case}

We assess our framework by upgrading a Java Moneta-based system across three synthetic repositories. This section outlines the repositories, explains the rationale for using synthetic data, and presents results comparing our framework’s upgrades to manual developer upgrades (“Gold standard”) and to the OpenHands agentic development tool \cite{wang2024openhands}.

\subsection{Synthetic Dataset}
The target dependency is a cloud-native, event-sourced microservice framework for Java, built on Spring Boot and widely used in industry. Upgrades often require multiple code changes across repositories, but not all updates are consistently applied. To ensure comprehensive evaluation, we created three synthetic repositories that encompass all necessary upgrade changes.
These repositories were manually upgraded to serve as the Gold standard for comparison. They reflect real-world industrial setups, including Java source files, yml configurations, and pom files.

\subsection{Results}

The framework is evaluated using the aforementioned three synthetic code repositories against the "Gold" (G) standard, each representing different upgrade scenarios for the framework. The repositories are designed to test the framework's ability to apply all necessary updates for specific version transitions: from versions 3.1 to 3.2, 3.2 to 3.3, and 3.3 to 3.4. We select a baseline (OpenHands) to benchmark against our \textbf{L}LM \textbf{A}gents for \textbf{D}ependency \textbf{U}pgrades (LADU). 

Table \ref{tab:results} compares the framework's performance with OpenHands + Claude 3.7 Sonnet and LADU + GPT-4o approaches. Key metrics include the number of files updated by each approach, the number of files in common with (G), the number of lines added or removed (A:R) by each approach, and common lines between the framework's output and (G). The  common file and line metrics are calculated as an exact match between the patch diffs. This is because LADU is run until unit tests pass, but the tests may not be comprehensive and do not cover deployment-related configurations in yaml files.

Table \ref{tab:runtime} presents a comparative analysis of the multi-agent LLM framework (LADU) against the OpenHands benchmark with regards to the execution metrics in order to assess efficiency. Number of steps, execution runtime in seconds, total tokens and cost in USD are reported.

\subsection{Discussion}

The results indicate that the proposed framework, while not perfectly matching the "gold" standard, demonstrates a significant ability to automate dependency upgrades with a reasonable degree of accuracy. The number of common files and lines suggests that the framework can effectively identify and implement necessary changes, although there is room for improvement in aligning with manual upgrades.

The first version upgrade was presented with the most comprehensive migration guide, which is reflected in the performance of both approaches in Table \ref{tab:results}. LADU is able to achieve comparable performance to the baseline whilst using GPT-4o and comprising of a much simpler codebase and structure. More specifically, with regards to recall, our framework is able to make more correct line changes with regards to the addition of relevant lines, but does not remove as many correct ones as OpenHands. 

In terms of precision, LADU attempts fewer changes, resulting in higher precision and reducing the risk of unwanted modifications. For example, in the 3.2 to 3.3 upgrade, LADU’s removals were more accurate (71.4\% vs. OpenHands’ 17.2\%).

Table \ref{tab:runtime} shows LADU requires far fewer steps (18 vs. 106 for OpenHands) and runs faster. It also uses significantly fewer tokens, especially in version 3.3, thanks to Meta-RAG’s codebase condensation. LADU’s token count includes summary generation, yet it still achieves results comparable to state-of-the-art benchmarks with greater efficiency.

\section{Conclusion}

This paper presents a novel framework consisting of multiple agents, including a Summary Agent, Control Agent, and Code Agent, which work together to update outdated library usages in codebases, while ensuring compatibility with new versions. The framework employs a change localization mechanism (Meta-RAG), which condenses the codebase, facilitating efficient information retrieval. The system was evaluated using synthetic code repositories, demonstrating its ability to automate upgrades with precision and efficiency compared to existing solutions. In the future, the evaluation can be expanded to include real-world codebases, and robust unit tests, to provide a more comprehensive assessment of the framework's effectiveness in industrial settings. Furthermore, integrating more advanced LLMs and exploring hybrid approaches might also improve performance. Overall, the framework represents a promising advancement in automated software maintenance, offering a scalable solution for managing Java dependency upgrades in large codebases.

\section*{Disclaimer}
This paper was prepared for informational purposes by the Artificial Intelligence Research group of JPMorgan Chase \& Co and its affiliates (“JP Morgan”), and is not a product of the Research Department of JP Morgan. JP Morgan makes no representation and warranty whatsoever and disclaims all liability, for the completeness, accuracy or reliability of the information contained herein. This document is not intended as investment research or investment advice, or a recommendation, offer or solicitation for the purchase or sale of any security, financial instrument, financial product or service, or to be used in any way for evaluating the merits of participating in any transaction, and shall not constitute a solicitation under any jurisdiction or to any person, if such solicitation under such jurisdiction or to such person would be unlawful.

\bibliographystyle{IEEEtran}
\bibliography{references}
\end{document}